\documentclass[a4paper]{article}
\usepackage[utf8]{inputenc}
\usepackage[english]{babel}

\usepackage{amsmath}
\usepackage{amssymb}
\usepackage{latexsym}

\usepackage{tikz}

\usepackage{amsthm}
\newtheorem{theorem}{Theorem}
\newtheorem{property}[theorem]{Property}

\usepackage{url}

\usepackage{xspace}
\usepackage{algorithm}
\usepackage{algorithmic}
\usepackage{color}

\usepackage{RR}

\RRetitle{Egomunities\\Exploring Socially Cohesive Person-based Communities}
\RRtitle{Egomunautés\\Exploration de communautés socialement cohésives et personne-centrées}
\RRdomaine{3}
\RRtheme{Réseaux et télécommunications}
\RRauthor{
  Adrien Friggeri
  \and
  Guillaume Chelius
  \and
  Eric Fleury
}
\URRhoneAlpes
\RRequipe{DNET}
\RRdate{11 February 2011}
\RRdater{18 February 2011}
\RRNo{7535}
\RRversion{2}

\RRmotcle {réseaux sociaux, réseaux complexes, graphes réels, détection de communautés, communautés recouvrantes, data mining, modélisation}
\RRkeyword{social networks, complex networks, real-world graphs, community detection, overlapping communities, data mining, modelisation}
\RRabstract{

The last years, there has been a great interest in detecting overlapping
communities in complex networks, which is understood as dense groups of nodes
featuring a low outbound density. To date, most methods used to uncover
overlapping communities stem from the field of disjoint community detection by
attempting to decompose the whole network into several possibly overlapping
groups of nodes. In this article we take an orthogonal approach by introducing a
novel point of view to the problem of overlapping communities, namely the concept
of egomunities, which are subjective communities centered around a given node,
more precisely inside its neighborhood. In order to construct those egomunities,
we propose a general metric on graphs, the cohesion, inspired by sociological
considerations. The cohesion quantifies the community-ness of one given set of
nodes, based on the notions of weak ties and triangles -- triplets of pairwise
connected nodes, instead of the classical view using only edge density. A set of
nodes has a high cohesion if it features a high density of triangles and
intersects few triangles with the rest of the network. We build upon the cohesion
to construct a heuristic algorithm which uncovers egomunities of a given node by
attempting to maximize their cohesion. We illustrate the pertinence of our method
with a detailed description of one person's egomunities among Facebook friends
and promising results from an ongoing large scale Facebook experiment. We finally
conclude by describing promising applications of egomunities such as information
inference and interest recommendations, and present a possible extension to
cohesion in the case of weighted networks.

}

\RRresume{ 

Ces dernières années, l'intérêt pour la détection de communautés recouvrante dans
les réseaux réels s'est intensifié. Celles ci sont des groupes de nœuds possédant
une forte densité interne et présentant une densité faible vers le reste du
réseau. À ce jour, la majorité des méthodes utilisées pour calculer de telles
communautés héritent de celles développées dans le domaine de la détection de
communautés disjointes, ou bien en étendant le concept de modularité à un
contexte recouvrant, ou bien en essayant de décomposer le réseau entier en
plusieurs sous ensembles éventuellement recouvrant. Dans ce rapport, nous
abordons la question de manière orthogonale en introduisant une mesure, la
cohésion, reposant sur des considérations sociologiques. La cohésion permet de
quantifier l'aspect communautaire d'un ensemble de nœuds à partir des notions de
triangles -- triplets de nœuds interconnectés -- et de liens faibles, au lieu de
la vision classique utilisant des arêtes. En substance, nous introduisons une
caractérisation numérique des communautés: des ensembles de nœuds possédant une
cohésion élevée. Nous présentons ensuite une nouvelle approche au problème des
communautés recouvrantes en introduisant le concept d'ego-munauté: des
communautés subjectives centrées sur un nœud donné, précisément incluses dans son
voisinage. Nous utilisons la cohésion pour élaborer un algorithme heuristique
construisant les ego-munautés d'un nœud en tentant de maximiser leur cohésion.
Finalement, nous présentons des résultats préliminaires, sous la forme d'une
description détaillé des ego-munautés d'amis Facebook d'une personne. Nous
concluons en décrivant des applications prometteuses des ego-munautés, par
exemple l'inférence d'information sur le sujet ou la recommandation de centre
d'intérêt, et présentons une extension possible à la cohésion dans le cas de
réseaux pondérés.

}

\newcommand{\neighbors}{\mathcal{N}}
\newcommand{\cohesion}{\mathcal{C}}
\newcommand{\tin}{\triangle_\text{in}}
\newcommand{\tout}{\triangle_\text{out}}
\newcommand{\degree}{\operatorname{deg}}

\begin{document}

\makeRR
%\maketitle

\section{Introduction}                                                            \label{sec:introduction}

\begin{sloppypar}

Although community detection has drawn tremendous amount of attention across the
sciences in the past decades, no formal consensus has been reached on the very
nature of what qualifies a community as such. In addition to the contributions of
sociology, several propositions have also emerged from the physics and computer
science communities~\cite{Castellano:2004p10,Flake:2002p24}. Despite the lack of
globally accepted analytical definition, all authors concur on the intuitive
notion that a community is a relatively dense group of nodes which somehow
features less links to the rest of the network. Unfortunately, this agreement
does not extend to the specific formal meanings of \emph{dense} and \emph{less
links}.
\end{sloppypar}

However, the past few years have witnessed a paradigm shift as the idea
of defining the nature of communities was progressively left aside. It has become
apparent, and widely accepted that it suffices to compare several sets of
communities and choose the \emph{best} obtained division -- relative to a given
metric -- in order to detect communities.

The metric the most used to that effect is Newman's
$Q$-modularity~\cite{Newman:2004p8}, which compares the density of links inside a
given community to what would be expected if edges where distributed randomly
across the network (null model). This method has proven to give sensible results on several
networks and gained traction in the \emph{communities} community. Since
maximizing the $Q$-modularity on general graphs is an established NP-hard
problem, several heuristics have been
proposed~\cite{louvain2008,Fortunato:2010p206,Sozio:2010p207}.

Most approaches were mainly focused on partitioning a network, leading to non
overlapping communities (each node belonging to one and only one group). In the
recent years, there has been a growing interest in the study of overlapping
communities; a distribution of the nodes across different groups which reflect
more precisely what one might expect intuitively, namely that a given node might
belong to different communities -- for example, in a social network, an
individual might simultaneously belong to a family, a friends group and
co-workers groups.

Due to the historical evolution of the field, to this day, most methods used to
detect overlapping communities are inspired by, or adapted from, existing
counterparts for disjoint community detection. If some of those methods take a
literal approach to the issue and are built upon extensions to the
modularity~\cite{Nepusz:2008p29,Shen:2009p30}, others have taken another path,
such as clique percolation~\cite{Palla:2005p32}. However, we assess that all
those methods aim at finding all communities in a network.

In this article, we propose to take a step back and focus on a specific type of a
user-centric communities, which we call \emph{egomunities}. Those egomunities are
overlapping communities contained in the neighborhood of one given node. In order
to detect those egomunities, we introduce a graph metric, the \emph{cohesion},
upon which we construct a heuristic algorithm. Drawing inspiration from well
established sociological results, the \emph{cohesion} is based on the notions of
weak ties and triangles -- triplets of pairwise connected node -- instead of the
classical view that uses edges to rate the \emph{communityness} of one given set
of nodes. Preliminary yet promising results from a large scale ongoing Facebook
experiment prove that the cohesion accurately captures the quality of a
community.

It is important to note that whereas the Q-modularity gives a score to a
partition of a network, the \emph{cohesion} serves as an intrinsic
characterization of a subgraph embedded in a network, independently of any other
subgraphs. As such, we propose a definition of a community, namely a set of nodes
with high cohesion. Moreover, even though the cohesion is a generic metric on
subgraphs, it was primarily conceived to characterize social communities. Its
inception relies on social considerations which formal extension to other types
of network is beyond the scope of this report. Therefore, we make no claim
towards or against its pertinence when used in the context of networks
representing non social data. 

This paper is organized as follows: in Section~\ref{sec:cohesion} we describe the
construction of a new metric, the \emph{cohesion}, to evaluate the
\emph{communityness} of a set of nodes. In Section~\ref{sec:egomunities} we
present a user-centric way of thinking about communities: \emph{ego}munities and
introduce an algorithm, relying on \emph{cohesion}, which computes those. In
Section~\ref{sec:results} we first present the egomunities of Facebook friends of
a test subject and evoke preliminary results of an ongoing large scale
experiment. Finally we highlight several applications and extensions.

\section{Cohesion}                                                                \label{sec:cohesion}
Before delving into technicalities and formal definitions, we consider important
to take a moment to reflect on the idea of community detection and highlight
inherent differences between the problems of disjoint versus overlapping
communities. We assess that the evaluation of the quality of a given set of
communities in a network mainly boils down to the two following questions:
\vspace*{-0.4cm}
\begin{itemize}
  \item \textbf{boundaries} : does the set of communities makes sense as a whole ? \vspace*{-0.3cm}
  \item \textbf{inside}    : is each community intrinsically sound ?
\end{itemize}
\vspace*{-0.4cm}
The main difference between disjoint and overlapping communities problems is that
in the latter a node can belong to several communities. Although seemingly
mundane, in the disjoint case this has for effect that ``\emph{belonging to the same
community}'' is an equivalence relation on nodes. As a consequence of this
relation's transitivity, the two aforementioned questions are deeply linked when
partitioning a network into communities.

This is actually the main idea behind $Q$-modularity, defined as follow: $Q =
\operatorname{Tr}\mathbf{e} - ||\mathbf{e}^2||$ where
$\operatorname{Tr}\mathbf{e}$ is the trace of the matrix $\mathbf{e}$, in which
$\mathbf{e}_{i,j}$ represents the density of links going from community $i$ to
community $j$. $Q$ increases when the communities are dense (\emph{i.e.} are
intrinsically sound) and decreases in presence of links between communities
(\emph{i.e.} when boundaries between communities are not well defined). In the
case of disjoint communities, optimizing the $Q$-modularity leads to a balance
between intrinsic and extrinsic qualities. Contrast this with the overlapping
problem, where those two questions are decoupled as one can modify one community
without affecting the others.

\subsection{The volatility of boundaries}                                         \label{sec:cohesion:volatility}
Of those two questions, within the scope of this paper, we evade the first one
for the most part, as we believe that methods to quantify the quality of a set of
communities should arise from choices adapted both to the analyzed data and to
the type of results one wishes to manipulate.

\begin{figure}[htb]
  \centering
  \begin{tikzpicture}[scale=1.2]
    \usetikzlibrary{patterns}
    \draw [color=black!30,pattern color=black!30,pattern=horizontal lines] (0,0.3) circle (0.3);
    \draw [color=black!30,pattern color=black!30,pattern=vertical lines] (0,-0.3) circle (0.3);
    
    \draw [color=black!30,pattern color=black!30,pattern=horizontal lines] (1,0.15) circle (0.3);
    \draw [color=black!30,pattern color=black!30,pattern=vertical lines]   (1,-0.15) circle (0.3);
    
    \draw [color=black!30,pattern color=black!30,pattern=horizontal lines] (2,0) circle (0.3);
    \draw [color=black!30,pattern color=black!30,pattern=vertical lines]   (2,0) circle (0.3);
    
  \end{tikzpicture}
  \caption{Three couple of cliques. On the left, there are clearly two communities, and on the right only one. The middle case is more of a gray area.}
  \label{fig:twocliquesboundaris}
\end{figure}
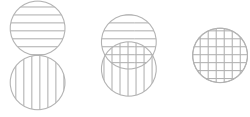

Consider for example two overlapping cliques. It seems reasonable to consider two
communities if the overlap is reduced to a single node, and one big community
when the intersection contains all nodes but one in each clique. The intermediate
case, however, is more of a gray area (Fig.~\ref{fig:twocliquesboundaris}). On
the one hand, it might be legitimate to consider only one community when two sets
of nodes feature a high enough overlap. In the field of network visualization,
for example, representing sets which intersect greatly each other could lead to
visual clutter, rendering the visual output unreadable. On the other hand, there
is a case for the opposite strategy, when the resulting communities should
be fine grained.

As such, the rating awarded to a set of communities should be tailored on a case
by case basis, in order to fit to the type of results which are sought.

\subsection{Focus on the inside}                                                  \label{sec:cohesion:focusontheinside}
It is possible to rate the quality of one given community embedded in a network,
independently from the rest of the network. The idea is to give a score to a
specific set of nodes describing wether the underlying topology is
\emph{community like}. In order to encompass the vastness of the definitions of
what a community is, we propose to build such a function, called \emph{cohesion},
upon the three following assumptions:
\vspace*{-0.4cm}
\begin{enumerate}
  \item the quality of a given community does not depend on the collateral existence of other communities;\vspace*{-0.2cm}
  \item nor it is affected by remote nodes of the network;\vspace*{-0.2cm}
  \item a community is a ``dense'' set of nodes in which information flows more easily than towards the rest of the network.
\end{enumerate}
The first point is a direct consequence of the previously exhibited dichotomy
between content and boundaries. The second one encapsulates an important and
often overlooked aspect of communities, namely their locality. A useful example
is to consider an individual and his communities; if two people meet in a remote
area of the network, this should not ripple up to him and affect his communities.

The last point is by far the most important in the construction of the cohesion.
The fundamental principle is linked to the commonly accepted notion that a
community is denser on the inside than towards the outside world, with a twist.

In \cite{Granovetter:1981p9}, Granovetter defines the notion of \emph{weak ties}
as edges connecting acquaintances, and argues that ``\emph{[\dots] social systems
lacking in weak ties will be fragmented and incoherent. New ideas will spread
slowly, scientific endeavors will be handicapped, and subgroups separated by
race, ethnicity, geography, or other characteristics will have difficulty
reaching a modus vivendi.}''. Furthermore, he states that a ``\emph{weak tie
[\dots] becomes not merely a trivial acquaintance tie but rather a crucial bridge
between the two densely knit clumps of close friends}''. And finally, he assesses
that \emph{local bridges} -- edges which do not belong to a triangle, that is a
set of three pairwise connected nodes -- are weak ties. For these reasons, we
consider that the structural backbone of communities does not lie solely in the
edges of the network, but rather in its triangles.

\begin{figure}[htb]
  \centering
  \newcommand{\fourclique}[1]{
    \draw [color=gray!100] (0,0) -- (1,0) -- (1,1) -- (0,1) -- (0, 0);
    \draw [color=gray!100] (1,0) -- (0,1) ;
    \draw [color=gray!100] (0,0) -- (1,1) ;
    \foreach \x/\y in {0/0,0/1,1/0,1/1}{
      \draw [color=gray!100,fill=#1](\x,\y) circle (0.2) ;
    }
  }
  \begin{tikzpicture}[scale=1.2]
    \draw [color=gray!100] (-1.75,-0.5) -- (0,0);
    \draw [color=gray!100] (-1.75, 0.5) -- (0,0);
    \draw [color=gray!100] (-0.75,-0.5) -- (0,0);
    \draw [color=gray!100] (-0.75, 0.5) -- (0,0);
    \begin{scope}[rotate=-45]
      \fourclique{gray!80}
    \end{scope}
    \begin{scope}[shift={(-1.75,-0.5)}]
      \fourclique{gray!10}
    \end{scope}
    %\foreach \k in {72,144,216,288}{
    %  \draw [very thick,color=gray!80] (0:1) -- (\k:1);
    %}
  \end{tikzpicture}
  \caption{Two communities featuring the same number of links towards the outside world but clearly different from a communityness standpoint.}
  \label{fig:triangles}
\end{figure}
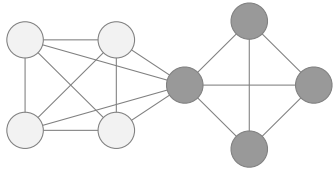

In Figure~\ref{fig:triangles}, two communities are represented in light and dark
gray. Both contain the same number of nodes and edges towards the rest of the
network. However, although it is sound to dismiss the lighter community as one of
bad quality -- as it is included in a larger clique -- the darker one is what one
would expect to be a community. Thus we are confronted with two sets of nodes,
featuring the same sizes, inner and outer densities, and yet one is a good
community and the other one is not. The difference between the two sets of nodes
appears when looking at triangles: the light set features six \emph{outbound}
triangles -- that is, triangles having an edge inside the community and a point
outside -- whereas the other set contains no such triangles.

Hence, we contend that the feature to consider when evaluating how well a
community's border is defined is not merely the presence of outbound edges, but
that of outbound triangles. Finally, we consider important to insist on the fact
that this metric does not describe how good is a set of communities but merely
the intrinsic quality of one community.

\subsection{Definition}                                                           \label{sec:cohesion:definition}
Given an undirected network $G=(V,E)$ and $S\in V$, we extend the notion of
neighborhood $\neighbors(G,u)$ of a node $u\in V$ to S, $\neighbors(G,
S)=\bigcup_{u \in S} \neighbors(G, u) \setminus S$.

\begin{figure}[htb]
  \centering
  \begin{tikzpicture}[scale=1.2]
    \draw [color=white,fill=gray!40] (0.5,0.5) circle (1.005);
    \draw [color=gray!100] (0,0) -- (0,1) -- (1,1) -- +(-30:1) -- (1,0) -- (0,0) -- (1,1) -- (1,0);
    \draw [color=gray!100,style=dashed] (0,0) -- (-1,0) -- (-1,1) -- (0,0);
    \draw [color=gray!100,fill=gray!20] (-1,1) circle (0.2);
    \draw [color=gray!100,fill=gray!20] (-1,0) circle (0.2);
    \draw [color=gray!100,fill=gray!20] (1.866,0.5) circle (0.2);
    \draw [color=gray!100,fill=gray!20] (0,0) circle (0.2);
    \draw [color=gray!100,fill=gray!20] (1,0) circle (0.2);
    \draw [color=gray!100,fill=gray!20] (1,1) circle (0.2);
    \draw [color=gray!100,fill=gray!20] (0,1) circle (0.2);
  \end{tikzpicture}
  \caption{Cohesion of a set of nodes (circled) in a network. $\tin=2$, $\tout=1$ (the dashed triangle is not taken into account as it has no edge in the set), therefore $\cohesion=\frac{1}{3}$.}
  \label{fig:cohesionexample}
\end{figure}
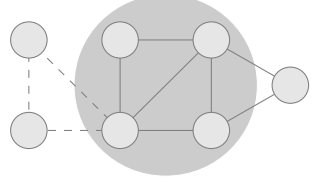

We first define two quantities, $\tin(G, S)$ which is the number of triangles of
$G$ contained in $S$ and $\tout(G, S)$ the number of triangles ``pointing
outwards'' — that is, triangles of $G$ having two nodes in $S$ and the third one
in $\neighbors(G, S)$. We then define the \emph{cohesion} $\cohesion$ of a subset
of nodes $S$ of a graph $G$:
\[
\cohesion(G, S) = \frac{\tin(G, S)}{{|S| \choose 3}}\frac{\tin(G, S)}{\tin(G, S)+\tout(G, S)}
\]
The first factor is the \emph{triangular density} of the community, while the
second one represents the proportion of triangles having a edge inside the
community which are wholly contained by said community. Intuitively, a community
has a high cohesion if it is dense in triangles \emph{and} it cuts few outbound
triangles. An example is given in Figure~\ref{fig:cohesionexample}. If there is
no ambiguity on the graph $G$, we will simplify the notation of $\cohesion(G, S)$
and note it: $\cohesion(S)$.

\subsection{Properties}                                                           \label{sec:cohesion:properties}
We assimilate the notions of \emph{weak tie} and \emph{local bridges}, and define
a weak tie as an edge which does not belong to any triangle. Let
$G_\triangle=(V,E_\triangle)$ be the graph obtained by removing all weak ties
from $G$.
\vspace*{-0.3cm}
\begin{property}
  For all $S\subseteq V$, $\cohesion(G, S) = \cohesion(G_\triangle, S)$.
  \begin{proof}
    When removing weak ties, no triangles are added or removed, thus
    $\tin(G,S)=\tin(G_\triangle, S)$ and $\tout(G,S)=\tout(G_\triangle, S)$.
    Therefore, $\cohesion(G, S) = \cohesion(G_\triangle, S)$.
  \end{proof}
\end{property}
\vspace*{-0.3cm}
This echoes the argument exposed in Section~\ref{sec:cohesion:focusontheinside}:
as weak ties serve only as links between communities, removing them from the
network does not affect communities quality.
\vspace*{-0.3cm}
\begin{property}
  \label{prop:disconnected}
  Let $S\subseteq V$ and $S^\prime\subseteq V$ be two disconnected sets of nodes $(\nexists e=(u,v) \in E \textnormal{ s.t. } u\in S \textnormal{ and } v\in S^\prime)$. If $\cohesion(S) <
  \cohesion(S\cup S^\prime)$ then $\cohesion(S^\prime) \geq \cohesion(S\cup
  S^\prime)$.
  \begin{proof}
    Suppose $\cohesion(S) < \cohesion(S\cup S^\prime)$ and $\cohesion(S^\prime) <
    \cohesion(S\cup S^\prime)$. From there it comes:
    \[
      \frac{\tin(S)^2}{{|S| \choose 3}} + \frac{\tin(S^\prime)^2}{{|S^\prime| \choose 3}} < \frac{(\tin(S)+\tin(s^\prime))^2}{{{|S| + |S^\prime|} \choose 3}}
    \]
    Given that $\forall a,b > 1, {a \choose 3}+{b \choose 3} < {{a+b} \choose 3}$,
    %\[{|s^\prime| \choose 3}^2\tin(s)^2 + {|s| \choose 3}^2\tin(s^\prime)^2 < 2{|s^\prime| \choose 3}{|s| \choose 3}\tin(s)\tin(s^\prime)\]
    \[
      \left({|S^\prime| \choose 3}\tin(S) - {|S| \choose 3}\tin(S^\prime)\right)^2 < 0
    \]
    Hence the contradiction.
  \end{proof}
\end{property}
\vspace*{-0.3cm}
As the cohesion of a group of nodes is a measure of its quality as a community,
it is understandable that adjoining a really good community to a lower quality
one might result in a group of nodes which is averagely good (consider for
example a huge clique and a poor set of nodes, the union might be more
\emph{community-ish} than the latter alone). Property~\ref{prop:disconnected} can
be understood the following way: if a community is disconnected, then one of its
connected component has a better cohesion than all connected component taken
altogether. As such, it makes sense to try to maximize the cohesion on connected
subgraphs. From now on, unless otherwise specified, we consider all cohesions on
a connected graph containing no weak ties.

We now present two analytical results. The first one is important as it exhibits
that sets of nodes conforming to the common definition of communities -- using
edge densities -- will obtain high cohesion. The second one shows that a large
clique does not shadow a smaller one if the overlap between the two is smaller
than a threshold depending on the size of the latter.

\begin{figure}[htb]
  \centering
  \includegraphics[width=0.8\linewidth]{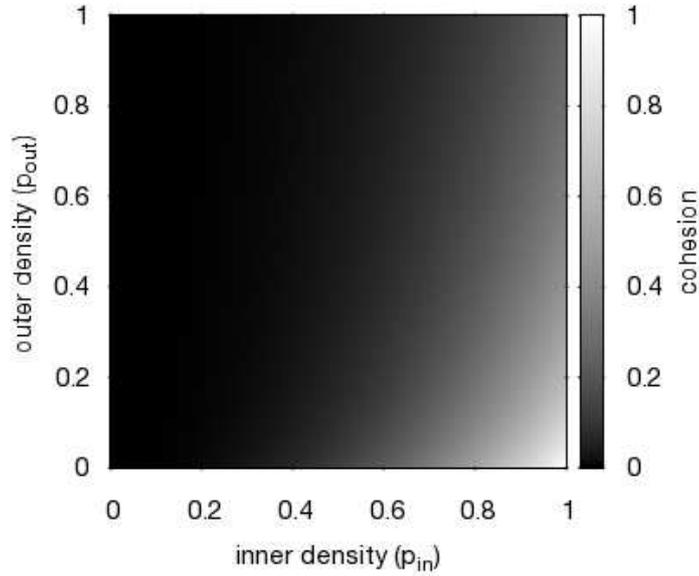}
  \caption{Cohesion of a set of 500 nodes connected to 500 external nodes as a function of inner and outer densities.}
  \label{fig:randomgraph}
\end{figure}

\noindent\textbf{Compatibility.} Let $S$ be a random network with an edge probability $p_{in}$ and suppose $S$ is
embedded in a network $G$, where an edge exist between each node of $S$ and each
node of $G$ with probability $p_{out}$. Then the cohesion of $S$ in $G$ is given
by:
\[\cohesion(S) = \frac{p_{in}^3}{1+\frac{3p_{out}|G|}{p_{in}(|S| -2)}}\]
Figure~\ref{fig:randomgraph} shows that when $S$ has a higher inner (resp. outer)
density, the cohesion increases (resp. decreases). This ensures that cohesion
remains compatible with the classical view on communities: it gives a higher
score to dense set of nodes featuring a low density to the outside world.

\begin{figure}[htb]
  \centering
  \includegraphics[width=0.8\linewidth]{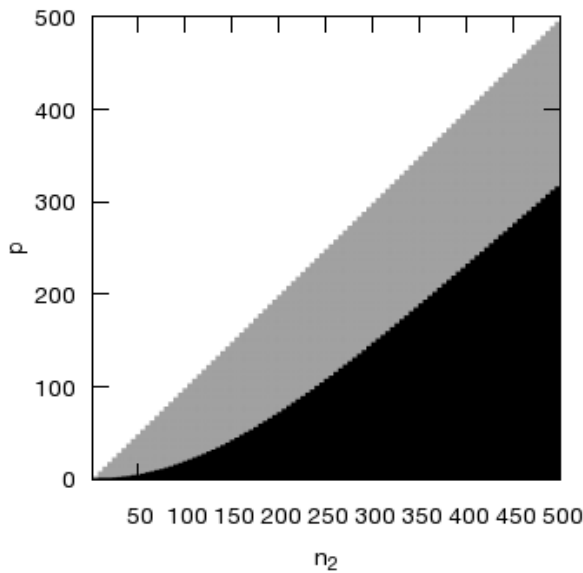}
  \caption{Regions where considering one community per clique (black) leads to a higher cohesion than considering only one big community (gray). $n_1 = 500$}.
  \label{fig:twocliques}
\end{figure}

\noindent\textbf{Non-shadowing.} We now consider a network containing two cliques $S_1$ and $S_2$ of size $n_1
\geq n_2$, having $p$ nodes in common. We have to the following cohesions :
\begin{eqnarray*}
  \cohesion(S_2) &=& \frac{1}{1+\frac{3(n_1-p)p(p-1)}{n_1(n_1-1)(n_1-2)}}\\
  \cohesion(S_1\cup S_2) &=& \frac{{n_1 \choose 3}+{n_2 \choose 3}-{p \choose 3}}{{{n_1+n_2-p} \choose 3}}
\end{eqnarray*}
In Figure~\ref{fig:twocliques}, we represent in black the region where
$\cohesion(S_2) \geq \cohesion(S_1\cup S_2)$. What this figure shows is that,
although $S_2$ might be much smaller than $S_1$, there is a threshold -- greater
than one common node -- under which $S_2$ has better cohesion than the whole
network, \emph{i.e.} a large clique does not always absorb a smaller one. This
ensures that cohesion does not suffer from resolution limit.

\section{Egomunities}                                                           \label{sec:egomunities}
\subsection{Interlude} 
As most recent works have focused on \emph{how} to detect communities, we deem
necessary to bring back the \emph{why} in the equation. It adds constraints to
the structure and type of communities one wishes to obtain: community detection,
in our opinion, has several purposes. First, as stated by Newman in his seminal
paper \cite{Newman:2004p8}, the ``\emph{ability to find and analyze such groups
can provide invaluable help in understanding and visualizing the structure of
networks}''. Hence, paraphrasing, detecting community is a way to \emph{simplify}
a complex topological structure in order to facilitate its visualization and
analysis.

If an algorithm produces an order of magnitude more than $n$ communities in a
network of size $n$ -- which incidentally cannot happen in the case of disjoint
communities but might be the case when considering overlapping sets of nodes --
the volume of data to deal with is not reduced but expanded and no simplification
occurs. This is striking when trying to visualize a network: the aim of
regrouping nodes into clusters is to reduce the clutter, not to pile up a great
deal of communities one on top of the other. However, graph compression is not
the only application of community detection.

Another possible use case lies in traits inference and social recommendation. The
past few years have witnessed the emergence of so-called \emph{online social
networks}, such as Facebook, LinkedIn, Twitter, etc. which have proven invaluable
as a source of data to study the structure of social interactions. The main
benefit of using such social networks is that they not only reproduce the
underlying social topology but add meta-information in the form of interests,
events, etc. They are however inherently limited by the fact that all information
they contain are subject to what the user reveals about himself. Therefore,
although the interpersonal links tend to be pretty exhaustive -- in terms of
\emph{who knows who} -- the information associated with each user is not. This
can be easily explained: whereas adding a connection to another user is a matter
of an instantaneous and simple click, entering one's centers of interest is time
consuming and is often done in an incremental manner.

However, as it is common knowledge that \emph{birds of a feather flock together},
it is possible to exploit the community structure of the network to infer what an
individual might be interested in. Consider for example a person and all their
acquaintances, if 1\% of those notified they liked going to a specific
restaurant, not much can be deduced. If however those 1\% represent 90\% of a
tight and coherent social community, chances are that the considered individual
has been to said restaurant. As such, community detection allows a refinement of
the social neighborhood in order to infer more precisely what might be relevant
to a given person, which has applications in terms of information discovery and
advertising.

In this user centric context, the relevance of a community set is defined by the
individual at the center in a subjective manner. In consequence looking for
communities at a global level -- the whole network -- might not be the best
approach. Consider for example a two spouses: both will have a \emph{family}
community, but might not include the same persons inside -- both will include
their children, their parents, maybe their in-laws, but when it comes to the
other spouses cousins their perception of what their family is might differ.

\subsection{Algorithm}
For the aforementioned reasons, we introduce the concept of \emph{egomunities},
namely person-based communities rooted in the subjective and local vision of the
network by a given node. In a manner of speaking, we attempt to bring a possibly
overlapping structure to the neighbors of the node. In this section we first
present a greedy algorithm which, given a network and a node, uncovers all
egomunities that this node belongs to. This is done by optimizing their cohesion
(Algorithm~\ref{algo:egomunities}). We then refine this algorithm by expanding
into several optimizations.

\begin{algorithm}[htb]
\caption{Greedy egomunities algorithm.}
\label{algo:egomunities}
\begin{algorithmic}
  \REQUIRE $G$ a graph, $u$ a node
  \STATE $E \gets \emptyset$
  \STATE $V \gets \neighbors(G,u)$
  \WHILE{$V \neq\emptyset$}
    \STATE $v \gets $ node with highest degree in $V$
    \STATE initialize the egomunity $\epsilon \gets \{u,v\}$ 
    \STATE $S \gets \{v^\prime\in\neighbors(G, \epsilon) / \cohesion(\epsilon\cup\{v^\prime\}) >\cohesion(\epsilon)\}$
    \WHILE{$S \neq\emptyset$}
      \STATE Add to $\epsilon$ the node $v\in S$ with the highest $\tin(\epsilon\cup\{v\})$, in case of ties, chose the node with the highest $\tout(\epsilon\cup\{v\})$
      \STATE $S \gets \{v^\prime\in\neighbors(G, \epsilon) / \cohesion(\epsilon\cup\{v^\prime\}) >\cohesion(\epsilon)\}$
    \ENDWHILE
    \STATE $V \gets V \setminus e$
    \STATE add $\epsilon$ to the set of egomunities $E \gets E \cup \{\epsilon\}$
  \ENDWHILE
  \RETURN the set of egomunities $E$
\end{algorithmic}
\end{algorithm}

Let $G$ be a network and $u$ a node of $G$, we focus on $u$'s neighborhood
$\neighbors(G, u)$ and discard the rest of the network. The core idea is to group
together neighbors in possibly overlapping egomunities, all containing $u$. To do
so, we initialize an egomunity by selecting a node $v_0 \in N$ to serve as
\emph{seed} -- thus the egomunity contains $u$ and $v_0$. From that point we
iterate and expand the egomunity by adding neighbors as long as it is possible to
increase the cohesion. If there are several nodes which addition increases the
cohesion, we choose to add the node $v$ which addition maximizes the number of
internal triangles $\tin$ -- and in the case more than one node satisfies this
condition, we select the one which maximizes the number of outbound triangles
$\tout$. Once no more node can be added to the egomunity, we start over by
selection the next seed from the sets of neighbors which haven't been assigned to
an egomunity and repeat the process until all neighbors are in at least one
egomunity.

The idea behind the algorithm is the following: each neighbor will be added, at
some point in time, to an egomunity. As such, it is possible to use any neighbor
as a seed; however, by choosing a node with a high degree in the neighbors
subgraph (that is, a node that forms a high number of triangles with the initial
node) as a seed, we create a set of nodes with a low $\tin$ and a high $\tout$.
The rationale behind the selection function in the greedy expansion phase is to
maximize $\tin$ as long as it results in a cohesion increase. We do not seek to
directly maximize the cohesion as this could lead to cases where one node is
selected because its addition decreases $\tout$ too much, thus limiting the
number of candidates at the next step. The exploratory phase can be seen as a
growth of an egomunity first by selecting the inner nodes and then only the
\emph{corners}.

For obvious reasons, it is costly to compute the cohesion at each step -- as it
would require at least to enumerate all triangles in one egomunity $\epsilon$, which
might be as high as ${|\epsilon| \choose 3}$. This gives a complexity of
$\mathcal{O}(n^3)$ if $|\mathcal{N}(u)|=n$ just to compute the cohesion. However,
it is possible to decrease the complexity by locally updating the cohesion when
adding a new node $v$ to $\epsilon$ :
$$  \cohesion(\epsilon\cup\{v\}) = \frac{(\tin(\epsilon) + I_v)^2}{({{|\epsilon|+1} \choose 3})(\tin(\epsilon) + \tout(\epsilon) + O_v)}$$
Where $I_v=\tin(\epsilon\cup\{v\}) - \tin(\epsilon)$ and $O_v=\tout(\epsilon\cup\{v\})-\tout(\epsilon)$)
are the number of inbound and outbound triangles which would be added to $\epsilon$
when including $v$. We now describe algorithm to add a node to an egomunity,
updating the cohesion and both quantities $I_v$ and $O_v$ for all impacted nodes.
It is important to remember that here, all egomunities contain one node in common
(the origin, $u$) and that because we restrict ourselves to the subgraph
containing only $u$ and its neighbors, $\neighbors(\{u,v\}) =
\neighbors(v)\setminus\{u\}$.

\begin{algorithm}[htb]
\caption{Updating when adding $v$ to an egomunity $\epsilon$.}
\label{algo:onlinecohesion}
\begin{algorithmic}
  \STATE $\tin \gets \tin + I_v$
  \STATE $\tout \gets \tout + O_v - I_v$
  \STATE $\epsilon \gets \epsilon \cup \{v\}$
  \FOR{$v^\prime \in \neighbors(G,v) \setminus \epsilon$}
    \STATE $n \gets \neighbors(G,v) \cap \neighbors(v^\prime)$
    \STATE $I_{v^\prime} \gets I_{v^\prime} + |n\cap \epsilon|$
    \STATE $O_{v^\prime} \gets O_{v^\prime} + |n\setminus \epsilon| - |n\cap \epsilon|$
  \ENDFOR
\end{algorithmic}
\end{algorithm}

\begin{figure}[htb]
  \centering
  \begin{tikzpicture}[scale=0.6]
    \draw [color=gray!20,fill=gray!20] (231:2.7) -- (159:2.8) .. controls +(18:2) and +(0,2) .. (309:2.7);
    \draw [dashed,color=gray!80] (159:2.8) .. controls +(18:2) and +(0,2) .. (309:2.7);
    \draw [color=gray!80,fill=gray!30] (231:2.7) .. controls +(0,2) and +(0,2) .. (309:2.7);
    \draw (18:2) -- (306:2) -- (162:2) -- (90:2) -- (306:2) -- (234:2) -- (18:2) -- (162:2) -- (234:2) ;
    \draw [color=gray!100,fill=gray!10] (306:2) circle (0.3) node {$v_0$} ;
    \draw [color=gray!100,fill=gray!10] (234:2) circle (0.3) node {$v_1$} ;
    \draw [color=gray!100,fill=gray!10] (162:2) circle (0.3) node {$v_2$} ;
    \draw [color=gray!100,fill=gray!10] (90:2)  circle (0.3) node {$v_3$};
    \draw [color=gray!100,fill=gray!10] (18:2)  circle (0.3) node {$v_4$};
  \end{tikzpicture}
  
  \caption{Updating the cohesion when adding $v_2$ to $\{v_0,v_1\}$, values for $I$ and $O$ are given inTable.~\ref{tab:onlinecohesion}.}
  \label{fig:onlinecohesion}
\end{figure}
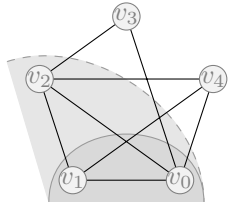

We first initialize, for all nodes $v$, $I_v = 0$ (there would be no triangles in
$e=\{u,v\}$) and $O_v = \degree(v)-1$ (all triangles having an edge $\{u,v\}$
would be cut, which is exactly the number of common neighbors to $u$ and $v$).
Then, each time a node is added to the egomunity, only the values pertaining to
its neighbors -- not included in $\epsilon$ -- need to be updated as described in
Algorithm.~\ref{algo:onlinecohesion} (Fig.~\ref{fig:onlinecohesion}).

\begin{table}[htb]
  \centering
  \begin{tabular}{|r|c|c|}
     \hline
     node  & ($I_v$, $O_v$) (before) & ($I_v$, $O_v$) (after) \\ \hline
     $v_2$ & (1, 4)                  & --                     \\ \hline 
     $v_3$ & (0, 1)                  & (1, 0)                 \\ \hline
     $v_4$ & (1, 3)                  & (3, 0)                 \\ \hline
  \end{tabular}
  \caption{\label{tab:onlinecohesion} Values for $I$ and $O$ before and after adding $v_2$ to $c=\{v_0,v_1\}$ as depicted on Fig.~\ref{fig:onlinecohesion}.}
\end{table}

\subsection{Two important heuristics}
As said earlier, the cohesion is conceived to judge the quality of a given
egomunity and not a set of egomunities, which is a totally different issue. The
algorithm as defined above generates overlapping egomunities in an independent
manner -- in regard to previous output. We assess that in some cases, obtaining
several groups of nodes which overlap too greatly might lead to irrelevant
results and propose a simple yet effective way of merging egomunities.

We define the overlap $\operatorname{overlap}(\epsilon_1, \epsilon_2) =
\frac{\epsilon_1 \cap \epsilon_2}{\operatorname{min}(|\epsilon_1|,
|\epsilon_2|)}$ and build an egomunity graph $G_E$ which nodes are egomunities,
and an edge $(\epsilon_1, \epsilon_2)$ exists if
$\operatorname{overlap}(\epsilon_1, \epsilon_2)$ is greater than a threshold
$o_\{\text{min}\}$. Although several approaches might be thought of in order to
carefully select which egomunities to merge (for example recursively computing
egomunities on $G_E$), we have observed that a less cumbersome yet resilient
method was to merge all egomunities pertaining to a same connected component in
$G_E$.

\begin{figure}[htb]
  \centering
  \begin{tikzpicture}[scale=0.6]
    \foreach \k in {0,60,120,180,240,300} {
      \foreach \l in {0,10,20,30,40}{
        \draw [color=gray!80] (\k:2) -- ++(160+\k+\l:2);      
      }
    }
    
    \draw [color=gray!80] (  0:2) -- ( 60:2);
    \draw [color=gray!80] (  0:2) -- (120:2);
    \draw [color=gray!80] (0:2) .. controls +(150:2.1) and +(30:2.1) .. (180:2);
    \draw [color=gray!80] ( 60:2) -- (120:2);
    \draw [color=gray!80] (180:2) -- (300:2);
    \draw [color=gray!80] (240:2) -- (300:2);
    \draw [color=gray!80] (240:2) -- (  0:2);
    \draw [color=gray!80,fill=gray!20] (0,0) circle (0.75) node{$c$};
    \draw [color=gray!100,fill=gray!10] (0:2)   circle (0.3) node{$v_0$};
    \draw [color=gray!100,fill=gray!10] (60:2)  circle (0.3) node{$v_1$};
    \draw [color=gray!100,fill=gray!10] (120:2) circle (0.3) node{$v_2$};
    \draw [color=gray!100,fill=gray!10] (180:2) circle (0.3) node{$v_3$};
    \draw [color=gray!100,fill=gray!10] (240:2) circle (0.3) node{$v_4$};
    \draw [color=gray!100,fill=gray!10] (300:2) circle (0.3) node{$v_5$};
  \end{tikzpicture}
  \caption{Network leading to overlapping egomunities.}
  \label{fig:costlygeneration}
\end{figure}
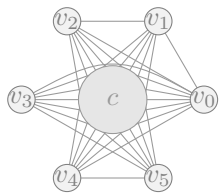

This merging step raises another issue: given the fact that some egomunities
might be merged, why bother and compute them separately in the first place? In
the worst case, given a neighborhood of $n$ nodes, the algorithm might output
$\frac{n}{2}$ egomunities containing each $1+\frac{n}{2}$ nodes. This is
illustrated in Fig.~\ref{fig:costlygeneration}, where up to 6 egomunities $\epsilon\cup
\{v_i\}$ might be generated only to be merged after hand. Given that computing
those distinct egomunities only is costly, we propose another heuristic in order
to reduce the useless calculations. After generating an egomunity, a last step is
done in which all nodes $v$ having a ratio $\frac{I_v}{O_v}$ greater than a given
threshold are added to that egomunity.

We deem important to point out the fact that this algorithm cannot be trivially
extended to compute overlapping communities on the whole network. An idea would
obviously be to generate egomunities for all nodes and consider the resulting set
of all egomunities as a set of communities for the network. This raises however
an issue, which is that, in a network $G=(V,E)$ of size $n$, containing $m$
edges, there is no \emph{a priori} reason that the egomunities for a node $u$
containing a node $v$ would be the same as the egomunities generated for $v$ and
containing $u$. Therefore, it would require to compare two by two the different
egomunities and decide whether they should be merged. As a node $u$ of degree
$d_u$ can generate at most $d_u$ communities, it would be necessary to compare
$\sum_{u,v\in V} d_u d_v$ egomunities, which is $\mathcal{O}(nm)$.

\section{Early results \& future works}                                    \label{sec:results}
In the previous sections we have defined a metric, the \emph{cohesion}, in order
to quantify the \emph{communityness} of a group of nodes and an algorithm which
produces egomunities of high cohesion for a given node. In order to validate both
the cohesion and the algorithm, we applied the latter to real world data. Through
the Facebook Graph API \cite{Facebook:2011}, it is possible to extract the social
neighborhood of a given individual. In this section we present preliminary
results which we obtained by computing egomunities for specific Facebook users,
first through a case study and then in the context of a large scale experiment.
Finally, we describe some possible applications and extensions.

\subsection{Case study}
We used our algorithm to compute for a few users their egomunities of Facebook
friends. We then interviewed those persons in order to determine if the
egomunities we obtained had a subjective meaning for them. In this section, we
present the results of one of those interviews.

\noindent\textbf{Egomunities.} The subject, a 32.5 year old male, had, at the time of the computation, 145
friends. Those friends were found to be distributed across 12 egomunities. 18
friends were not present in any egomunity (for example, friends having no
friends in common with the subject), 94 were in only one egomunity, 26 in two
egomunities, 3 in three and 4 in four different egomunities.
Table~\ref{tab:vcomms} lists those egomunities along with there size and
cohesion. A quick interview of the subject was conducted in order to figure if
each group had a social meaning to them and if so, how they would describe it.
All but one group echoed a specific part of the subject's life. However, it is
important to note that those egomunities only reflect the underlying Facebook
network, which may be incomplete and differ from the real world social network.

\begin{table}[htb]
  \centering
  \begin{tabular}{|c|r|r|r|}
     \hline
     description                           & size & cohesion \\ \hline
     higher education                      &   7  &   0.64   \\ \hline
     research (france)                     &   5  &   0.61   \\ \hline
     elementary school                     &   8  &   0.49   \\ \hline
     friends in Brazil                     &  10  &   0.38   \\ \hline
     circle of friend                      &  31  &   0.25   \\ \hline
     family                                &  10  &   0.22   \\ \hline
     brazilian dancers/musicians 1         &  11  &   0.19   \\ \hline
     capoeira                              &  13  &   0.17   \\ \hline
     dance                                 &  22  &   0.14   \\ \hline
     group of close friends                &   5  &   0.11   \\ \hline
     brazilian dancers/musicians 2         &   9  &   0.09   \\ \hline
     vague (mostly dance related)          &  52  &   0.07   \\ \hline
  \end{tabular}
  \caption{\label{tab:vcomms} Egomunities ordered by cohesion. A short description of what people in the same group have in common is given.}
\end{table}

\begin{figure}[htb]
  \centering
  \includegraphics[width=\linewidth]{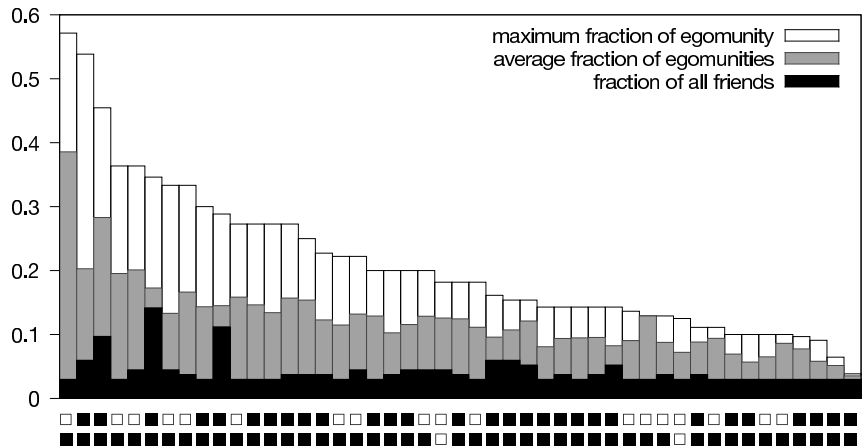}
  \caption{Proportions of all friends sharing a \emph{like}, and average, maximum proportions on egomunities.}
  \label{fig:likes}
\end{figure}

\noindent\textbf{Traits inference.} Ongoing work focuses on traits inference
based on egomunity structure. For example, it is possible to access some
information about a user on Facebook such as center of interest. We outline here
an idea on how to exploit egomunities to mine more accurate information on a
given individual. It is possible to gain some insight on one's centers of
interest by observing their \emph{likes}. Those are center of interests which
were specified by the users and might be shared between them. In
Figure~\ref{fig:likes}, we have extracted a subset of those interests satisfying
the following criteria: \emph{(i)} having been \emph{liked} by at least 4 of the
subject's friends and \emph{(ii)} absent from the user's \emph{likes}. Each
column represents a particular interest and we plotted, for each of those: the
proportion of all friends having this interest, the average proportion of friends
sharing this interest in the subject's egomunities where the interest appears and
the maximum proportion across egomunities. The abscissa also features two
squares. On the top row, a full (resp. empty) square indicates that the subject
was aware (resp. not aware) of the existence of the interest. The bottom row
indicates whether it might be of interest to the subject. In this particular
case, more than 95\% of the \emph{likes} were relevant to the subject (with no a
priori knowledge on his centers of interests): 61.7\% were already known but had
not been specified on Facebook, and 34\% were new \emph{likes} which were of
interest to him. Only two \emph{likes} were of no interest to the subject, and it
is notable that those do not feature a high maximum proportion across all
egomunities. It is moreover interesting to observe that out of the 8 interests
having the highest maximal proportion in an egomunity, the majority was unknown
to the subject despite being of interest to him after hand.

\subsection{The Fellows Experiment}
Building on the Facebook, API we have launched Fellows \cite{Fellows:2011}, a
large scale experiment on Facebook in which users are able to compute their
egomunities and rate them. The data we are collecting from this ongoing
experiment will allow us to statistically confront the cohesion model to
individual perception of egomunities. In this section we present preliminary
results obtained the data collected at the point of writing.

The participants were presented with an application which, once connected to
Facebook, analyzes their social neighborhood and presents them with their
egomunities computed by our algorithm. They are then asked to give a numerical
rating between $1$ and $4$, answering the question ``\emph{would you say that
this list of friends forms a group for you?}''. We collect all egomunities with
their cohesion and size. As the participant could stop rating at any time, some
egomunities have no ratings.

\begin{figure}[htb]
  \centering
  \includegraphics[width=0.9\linewidth]{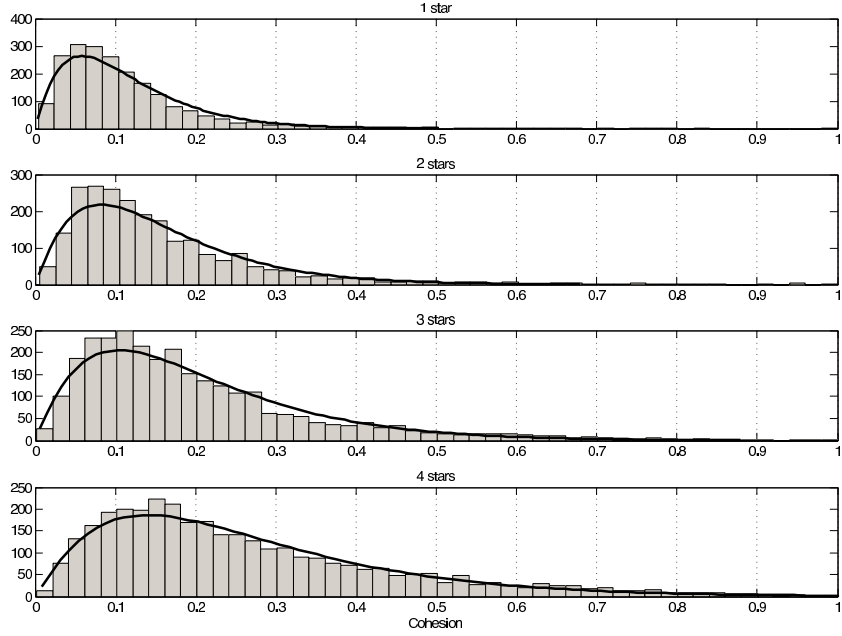}
  \caption{Density of cohesion for egomunities of rating 1, 2, 3 and 4.}
  \label{fig:cohesionDistribs}
\end{figure}

\begin{figure}[htb]
  \centering
  \includegraphics[width=\linewidth]{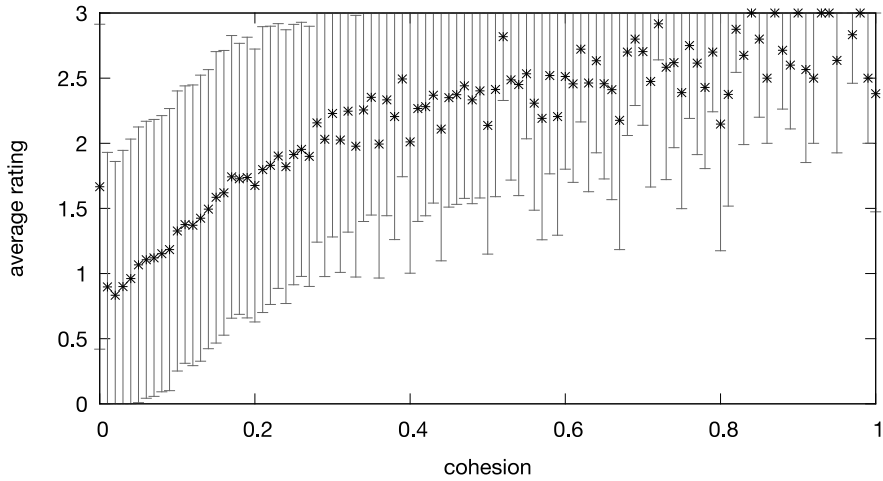}
  \caption{Average rating \emph{vs.} cohesion.}
  \label{fig:cohesionRatings}
\end{figure}

The result we present here are extracted from data collected from 980
participants, which totaled 22,697 egomunities and of those, 14,634 were rated.
On Figure~\ref{fig:cohesionDistribs} we group the rated communities by rating and
represent the distributions of cohesion for each rating, this shows that on
average egomunities with higher rating feature a higher cohesion. Conversely, in
Figure~\ref{fig:cohesionRatings}, we plot the average rating obtained by
egomunities grouped by cohesion slices of width $1/100^{\textnormal{th}}$ ; in turn, this
shows that on average, egomunities with higher cohesion tend to obtain higher
ratings. Hence we conclude that \emph{cohesion} provides an accurate
quantification of an egomunity's quality, as perceived by its original node.

Considering the fact that gathering all \emph{likes} for all users is intrusive
and might put off some participants, we decided to focus on the inference of
other traits, such as their age. Due to the presence of family, older co-workers,
younger siblings, etc. the neighborhood of a given person features age
disparities. However, all egomunities do not suffer from such heterogeneity:
although someone may have some friends of disparate ages, it is likely that at
least one of their egomunity features a low age variability (for example, an
egomunity of classmates). Our idea is to exploit this fact to pinpoint more
accurately the user's age.

\begin{figure}[htb]
  \centering
  \includegraphics[width=\linewidth]{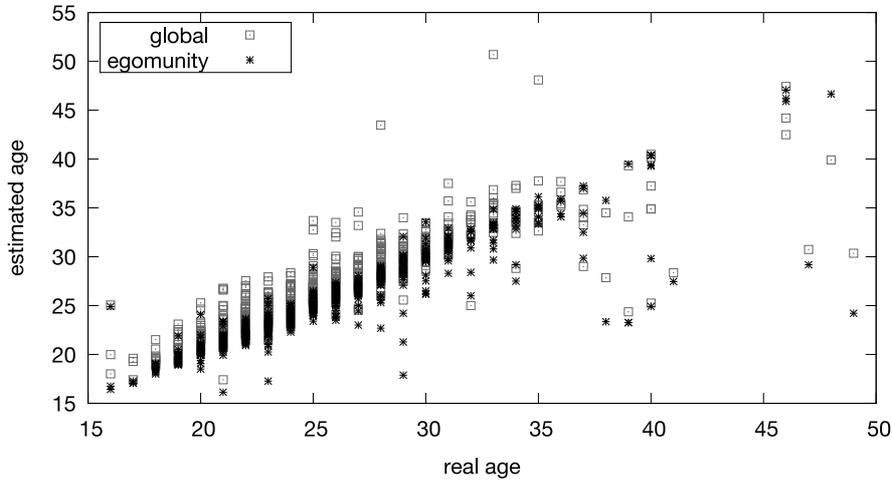}
  \caption{Subject age \emph{vs.} estimated age on all friends and most homogeneous egomunity.}
  \label{fig:ages}
\end{figure}

From now on, we only take into account egomunities of size greater than 10 and of
which the age standard deviation is less than 2.5 years (70.24\% participants
feature at least one such egomunity). Let $a$ be the vector where $a_i$ is the
age of the $i^\textrm{th}$ participant. We then define the \emph{globally
estimated age} $g_i$ as the average of all the $i^\textrm{th}$ participant's
friends, the \emph{egomunity based age} $e_i$ as the average age of the members
of the participant's egomunity featuring the lowest relative standard deviation.
Figure~\ref{fig:ages} shows $g$ and $e$ in relation to $a$. Given that both
quantities are correlated to $a$ (Pearson correlation coefficients: $r_{a,g} =
0.859$ and $r_{a,e} = 0.894$), we assess that they can be used to infer real
ages. However, the bias when considering all friends is of $+1.27$ years whereas
it is only $-0.296$ years when using only the less variable egomunity. Both
estimations feature similar variability ($\sigma_g=2.9$ and $\sigma_e=2.3$), but
the average absolute error is of 1.96 years when using $g$ whereas it is of 0.938
years in the other case. We conclude that the egomunity based age leads to a more
accurate estimation of the participant's age.

We conclude that although it is possible to infer a person's age using all their
friends, it is even more precise to do so using only a portion of their friends,
pertaining to the egomunity featuring the lowest relative variability.

\subsection{Extension to weighted networks}
Besides traits inference, future works will also focus on the evaluation of
weighted cohesion to quantify the quality of weighted social communities. In a
simple unweighted model of social networks, when two people know each other,
their is a link between them. In real life however, things are more subtle, as
the relationships are not quite as binary: two close friends have a stronger bond
than two acquaintances. In this case, weighted networks are a better model to
describe social connections, this is why we deem necessary to introduce an
extension of the cohesion to those networks.

The definition of the cohesion can, as a matter of fact, be extended to take the
weights on edges into account. We make the assumption on the underlying network
that all weights on edges are normalized between 0 and 1. A weight $W(u,v)=0$
meaning that there is no edge (or a null edge) between $u$ and $v$, and a weight
of 1 indicating a strong tie. We define the weight of a triplet of nodes as the
product of its edges weights $W(u,v,w) = W(u,v)W(u,w)W(v,w)$. It then comes 
that a triplet has a strictly positive weight if and only if it is a triangle. We
then define inbound and outbound weights of triangles and finally extend the
cohesion.
\begin{eqnarray*}
  \tin^w(S)      &=& \frac{1}{3}\sum_{(u,v,w) \in S^3} W(u,v,w)\\
  \tout^w(S)     &=& \frac{1}{2}\sum_{u\not\in S, (v,w)\in S^2} W(u,v,w)\\
  \cohesion^w(S) &=& \frac{\tin^w(S)}{{|S| \choose 3}}\frac{\tin^w(S)}{\tin^w(S)+\tout^w(S)}
\end{eqnarray*}

\section{Conclusion}

In this article we have presented a novel take to uncovering overlapping
communities. Our approach lies in the use of egomunities: a person-based point of
view of their neighbors' communities. To that effect, we define a metric, the
\emph{cohesion}, to quantify the intrinsic communityness of any subset of nodes
of a network. We used the cohesion to design an algorithm which constructs
egomunities. We applied this algorithm on data extracted from Facebook, both in
the form of case studies and through a large scale ongoing experiment called
Fellows, in which users are presented with their egomunities and are asked to
rate them according to their own perception. The experiment provides us with data
which already tend to validate the accuracy of cohesion as a community quality
measure. Moreover, preliminary results are promising, as we were able to exhibit
that the use of egomunities can lead to the construction of efficient estimators
for several personal traits. Future work will rely on data collected during the
Fellows\footnote{\url{http://fellows-exp.com/}} experiment to further our study
on traits inference. Using the weighted cohesion, we will also investigate the
influence of weights on egomunity detection.

\newpage
\bibliographystyle{abbrv}
\bibliography{biblio}
\end{document}